\documentclass[usenatbib]{mnras}

\usepackage{newtxtext,newtxmath}
\usepackage[T1]{fontenc}

\DeclareRobustCommand{\VAN}[3]{#2}
\let\VANthebibliography\thebibliography
\def\thebibliography{\DeclareRobustCommand{\VAN}[3]{##3}\VANthebibliography}

\usepackage{subcaption}
\usepackage{graphicx}	% Including figure files
\usepackage{amsmath}	% Advanced maths commands
\title[PBH lunar craters]{Crater Morphology of Primordial Black Hole Impacts}

\author[A. Yalinewich and M. Caplan]{
Almog Yalinewich$^{1}$\thanks{E-mail: almog.yalin@gmail.com}
and Matthew E. Caplan,$^{2}$
\\
$^{1}$Canadian Institute for Theoretical Astrophysics, 60 St. George St., Toronto, ON M5S 3H8, Canada\\
$^{2}$Department of Physics, Illinois State University, Normal,IL 61790 USA
}

\date{Accepted XXX. Received YYY; in original form ZZZ}

\pubyear{2020}

\begin{document}
\label{firstpage}
\pagerange{\pageref{firstpage}--\pageref{lastpage}}
\maketitle

\begin{abstract}
In this work we propose a novel campaign for constraining relativistically compact MACHO dark matter, such as primordial black holes (PBHs), using the moon as a detector. PBHs of about $10^{19} \textrm{ g}$ to $10^{22} \textrm{ g}$ may be sufficiently abundant to have collided with the moon in the history of the solar system. 
We show that the crater profiles of a PBH collision differ from traditional impactors and may be detectable in high resolution lunar surface scans now available. Any candidates may serve as sites for in situ measurements to identify high pressure phases of matter which may have formed near the PBH during the encounter.  
While we primarily consider PBH dark matter, the discussion generalises to the entire family of MACHO candidates with relativistic compactness. Moreover, we focus on the Moon since it has been studied well, but the same principles can be applied to other rocky bodies in our solar system without an atmosphere.

\end{abstract}


%%%%%%%%%%%%%%%%%%%%%%%%%%%%%%%%%%%%%%%%%%%%%%%%%%

%%%%%%%%%%%%%%%%% BODY OF PAPER %%%%%%%%%%%%%%%%%%

\section{Introduction}

One long standing question in astrophysics and cosmology is the nature of dark matter \cite[e.g.][ and reference therein]{Bertone2018HistoryMatter}. Candidates for dark matter particles can be divided into two categories: microscopic and macroscopic, where the latter are often referred to as MACHOs (massive compact halo objects). It has been suggested that dark matter is composed (at least in part) of black holes formed from density fluctuations in the early universe \citep{Carr2016PrimordialMatter}. Such objects are called primordial black holes (PBHs). Since the idea has been proposed, different observations have excluded certain mass ranges for these primordial black holes \citep{Carr2020ConstraintsHoles, Carr2020PrimordialDevelopments}. Currently, there are several poorly constrained mass windows, notably between $10^{16} \, \rm g$ and $10^{24} \, \rm g$, where the constraints are weak enough that PBHs could be a large fraction of the dark matter, though the exact mass windows are dependent on the PBH mass spectrum. While several constraints have been reported in this mass range some have been contested \citep{Katz2018FemtolensingRevisited}, which motivates the present work to develop a new independent constraint in this asteroid-mass to sublunar-mass window.
 %Currently, there is a window of mass ranges, between $10^{16} \, \rm g$ and $10^{24} \, \rm g$ that is allowed (i.e. unconstrained by observations). We note that a previous work claimed to have excluded this range, but a later study contested this claim \citep{Katz2018FemtolensingRevisited}. 

In this study we consider the possibility that a primordial black hole collided with the moon in the past. As we shall show later, the  craters from such collision will have a shape markedly different from an ordinary meteor impact. This feature can readily be detected by surface images and radar scans.

The idea of a collision between a primordial black hole (or another compact object) with planetary bodies has been considered by various authors in the past.  \cite{Rafelski2013CompactImpactors} have considered in some detail the physics of compact ultra-dense matter colliding with solar system bodies. They found that relativistically compact matter will generally completely penetrate the body forming both `entrance' and `exit' wounds on a hyperbolic trajectory. Penetration is generally supersonic which may produce a seismic signal \citep{Khriplovich2008CanEarth} which has been searched for \citep{ Herrin2006SeismicNuggets, Banerdt2007UsingNuggets, Horowitz2020GravimeterEarth}. Accretion onto a high velocity PBH `entering' the earth has been proposed as an explanation for the Tunguska event \citep{Jackson1973WasHole}. `Exit' events have meanwhile been compared to volcanic events, transporting crust material high into the atmosphere \citep{Rafelski2013CompactImpactors, Rafelski2013CompactSystem}. While surface processes on the earth complicate efforts to establish historic events as PBH encounters, the Moon and other solar system bodies without active surface processes hold a record of several billion years of impacts in their surface geology. In this work we develop the theory of cratering dynamics in PBH encounters with solar bodies and describe a possible observational campaign to constrain low mass MACHO dark matter. 

While we generally consider PBHs interacting with the Moon in this work, our discussion generalises to many solar system bodies and ultradense dark matter candidates.

The structure of the paper is as follows. In section \ref{sec:cratering} we present analytic calculations of the shock wave trajectories in the case of a regular cratering event and in the case of a PBH impact. Using the shock trajectories, we calculate how the ejecta will be distributed outside the crater, and show that each one of these cases will yield a unique ejecta distribution. In section \ref{sec:sims} we present numeric hydrodynamic simulations that verify our analytic results. Finally, in section \ref{sec:conclusion} we discuss the results and the implications for PBHs as dark matter particles.

%observational campaigns to search for high resolution lunar 

%Primordial  black holes (PBH) as a MACHO candidate are constrained at the lower limit, $~10^{18}$ g, by evaporation to Hawking radiation, though other non PBH MACHOs may be possible. These include baryonic matter near nuclear density such as `strangelets' or agglomerations of dark sector particles. 

\section{Cratering} \label{sec:cratering}

\subsection{Classical Cratering}

We begin by describing classical craters, which form as a result of meteor impacts or surface explosions. In the former, an impactor with mass $m_i$ collides with ground at some velocity $v_i$. For simplicity, we assume that both the impactor and the ground have the same density $\rho_0$. As a result of the collision, a shock wave emerges from the impact site and travels into the soil. As the shock wave expands, it goes through few distinct phases. In the first phase, while the distance travelled by the shock is much smaller than the radius of the impactor, the shock velocity is roughly uniform and roughly equal to $v_i$. In the second stage, the shock begins to decelerate as it sweeps up more mass. As the shock expands, it loses information about the impactor's incidence angle, and converges to the same universal solution \citep{Yalinewich2020SelfCollisions}. For this reason, we can just consider head on  collisions. In the case of hypervelocity impacts, the velocity decelerates according to some power law in the swept up mass $v \approx v_i \left(m/m_i\right)^{-\beta}$ where $\beta$ is some constant. Energy and momentum conservation provide limits on the value of this power law index \citep{Zeldovich1956ThePressure} $1>\beta>1/2$. The value of $\beta$ depends on the equation of state of the soil \citep{Housen1983CraterAnalysis., Yalinewich2020SelfCollisions}, but is typically close to 2/3. As the shock excavates a crater basin, material is expelled out of the mouth of the basin. A fluid element emerges from the basin with roughly the same velocity with which it was shocked. The expelled material eventually settles outside the rim of the crater, forming the so called ejecta blanket \citep{Osinski2011ImpactPlanets}. The process by which an ejecta blanket forms is referred to as ballistic sedimentation. The final radius of debris expelled at a velocity $v$ is $r \approx v^2/g$, where $g$ is the gravitational acceleration. We can use the velocity profile of the shock to calculate the surface density profile of the ejecta blanket
\begin{equation}
    \sigma \approx \frac{m}{r^2} \approx r^{-2-1/2\beta} \, .
\end{equation}
Energy and momentum conservation yield limits on the slope of the surface density distribution $-2.5>d \ln \sigma / d \ln r>-3$. If we assume a constant density for the settled ejecta, then the thickness of the ejecta blanket scales the same way as the surface density. Most ejecta blankets around craters on the moon are within this range \citep{Zhu2015NumericalMoon, Zhu2017EffectsDistribution, Raggio2016ADVANCEMENTSCRATERS}. In contrast, ejecta blankets from chemical and nuclear explosions on Earth are slightly steeper $-3>d \ln \sigma / d \ln r>-4$, due to air drag \citep{Carlson1965DistributionSoils}.

Eventually, the shock stops and the crater reaches its final size. There are two criteria that can determine the final size of the crater (which will always be the smaller of the two). One possibility is that that the shock stops when the shock pressure drops below the elastic modulus of the soil $Y$, so the final shock radius is
\begin{equation}
    R \approx R_i \left(\frac{\rho_0 v_i^2}{Y}\right)^{1/6\beta}
\end{equation}
where $R_i \approx \left(m_i/\rho_0\right)^{1/3}$ is the radius of the impactor. The second possibility is that the shock stops when the shock pressure drops below the hydrostatic pressure $g \rho R$, and so
\begin{equation}
    R \approx R_i \left(\frac{v_i^2}{g R_i}\right)^{1/\left(6\beta+1\right)} \, .
\end{equation}
In the case of the lunar regolith, this is the deciding condition.

\subsection{Craters from PBHs}

The density of PBHs is much higher than that of meteors, so instead of transferring all of their energy to the target at the impact site, the impactor just passes through and deposits roughly the same amount of energy per unit length along its trajectory inside the target. To understand how the shock wave will behave in this case, one can think about this problem as a series of point explosions, where the energy of each explosion scales with its depth $E \propto x$. If the mass of the black hole is $m_i$, then the effective cross section is determined by the Bondi radius $R_b \approx G m_i /v_i^2$, and so $E \approx x R_b^2 v_i^2 \rho$. For a uniform density soil, the mass in a sphere of radius $x$ around the hot spot is $m \approx \rho_0 x^3$. Hence, $v \approx v_i R_b \rho_0^{1/3} m^{-1/3}$. We thus obtain a new power law relation between the shock velocity and the swept up mass, which is shallower than that obtained for classical craters, i.e. $\beta = 1/3$. Using the same reasoning as in the previous section we find the slopes of ejecta blanket surface density profile
\begin{equation}
    \frac{d \ln \sigma}{d \ln r} = -3.5 \, .
\end{equation}
To find the size of the crater, we need to find the depth at which when the shock reaches the surface, the shock velocity is comparable to the free fall velocity from the top of the basin to the bottom. The final crater radius is therefore
% https://www.wolframalpha.com/input/?i=%28%28G*%281e19+g%29%2F%28200+km%2Fs%29%29%5E2%2F%28gravity+on+the+moon%29%29%5E%281%2F3%29
\begin{equation}
    R \approx \left(\frac{v_i^2 R_b^2}{g}\right)^{1/3} \approx 1.9 \left(\frac{v_i}{200 \, \rm \frac{km}{s}}\right)^{2/3} \left(\frac{m_i}{10^{19} \, \rm g}\right)^{2/3} \left(\frac{g}{2 \, \rm \frac{m}{s^2}}\right)^{-1/3} \, \rm m \, .
\end{equation}
Craters of this size could be detected, for example, with the Lunar Reconnaissance Orbiter, which has a resolution of about 1 metre
\citep{Robinson2010LunarOverview}. Hence, any interaction with black holes more massive than $10^{19} \, \rm g$ could, in principle, be detected. The number of such collisions in the lifetime of the solar system is
%https://www.wolframalpha.com/input/?i=%288e-25+g%2Fcc%29%2F%281e19+gram%29*%28200+km%2Fs%29*%28radius+of+the+moon%29%5E2*%284+Gyr%29*4*pi
\begin{equation}
    N \approx 0.1 \left(\frac{\rho_{DM}}{8 \cdot 10^{-25} \, \rm g/cm^3}\right) \left(\frac{m_i}{10^{19} \, \rm g}\right)^{-1} \left(\frac{v_i}{200 \, \rm km/s}\right)
\end{equation}
where $\rho_{DM}$ is the density of dark matter. While this does not necessarily guarantee a detectable encounter between the moon and a PBH in the history of the solar system, it is of sufficiently high probability to be interesting. Furthermore, if the solar system has passed through regions of higher dark matter density in the past the expected number of events could be higher.

\section{Numerical Simulations} \label{sec:sims}

\begin{figure*}
    \centering
    \includegraphics[width=0.6\textwidth]{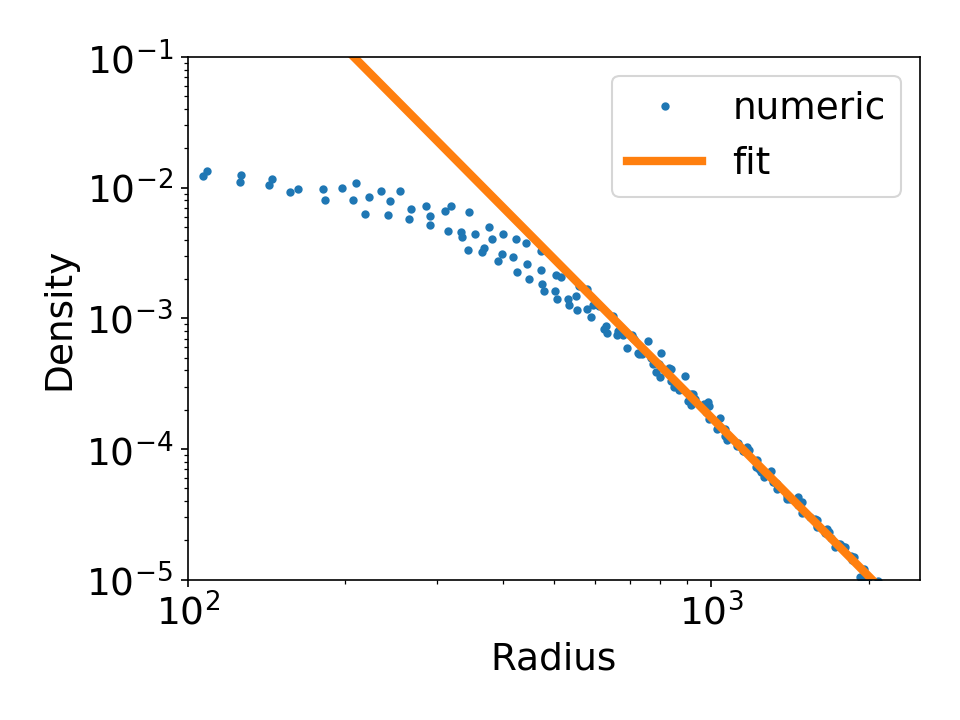}
    \caption{Radial density profile for a normal cratering event, obtained from numerical simulations. The radius is given in units of the impactor radius, and the density in units of impactor (and target) density. The slope of the power law fit is -4.1, whereas the analytic prediction is -4.5.}
    \label{fig:reference_pbh}
\end{figure*}

In order to verify the analytic results derived in the previous section we performed hydrodynamic simulations using the moving mesh code RICH \citep{Yalinewich2015Rich:Mesh}. In the first simulation we collide a cold sphere with a cold half space, both filled with ideal gas with the same density. This simulation is supposed to represent a normal cratering event. We chose the adiabatic index to be $\gamma=1.4$, since this value best describes the high pressure shock behaviour in silica \citep{Yalinewich2019AtmosphericImpacts}. Since we want to resolve the impactor and also follow the shock wave to distances much larger than the impactor radius, we arranged 20,000 mesh generating points in a non-uniform way inside the computational domain. In the vicinity of the impactor the resolution is 0.1 of the radius of the impactor, but at larger radii the cells are larger such that the ratio between the cell size and radius is roughly constant around 0.1. With this distribution of points, we were able to extend the boundaries of the computational domain to $10^6$ impactor radii. To simplify the simulation, we assume a head on collision, and run the simulation on a 2D grid using cylindrical symmetry. The density of both the impactor and the target is set to 1 while the density in the rest of the computational domain is $10^{-9}$, the velocity with which the impactor travels toward the target is set to 1 and the ambient pressure before impact is $10^{-9}$ everywhere. We ran the simulation to time $10^7$, where the unit of time here is the time it takes the impactor to traverse its own radius. The simulation does not include gravity, so the debris expands ballistically from the impact site. In figure \ref{fig:reference_pbh} we plotted the density of the ejecta as a function of distance from the impact site. Using the results of the previous section we can estimate the slope of this profile analytically. We saw that $v \propto m^{-\beta}$. At distances much larger than the radius of the impactor, the velocity profile is roughly homologous $v \approx d/t$, where $d$ is the distance from the impact site and $t$ is the time since the impact. Combining these relations with the definition of density $m \approx \rho d^3$ we find a scaling relation for the density profile $\rho \propto r^{-3 -1/\beta}$. For $\beta = 2/3$ we obtain $d \ln \rho / d \ln r = -4.5$, which is close to the value obtained in the numeric simulation $d \ln \rho / d \ln r = -4.1$.

\begin{figure*}
    \centering
    \includegraphics[width=0.6\textwidth]{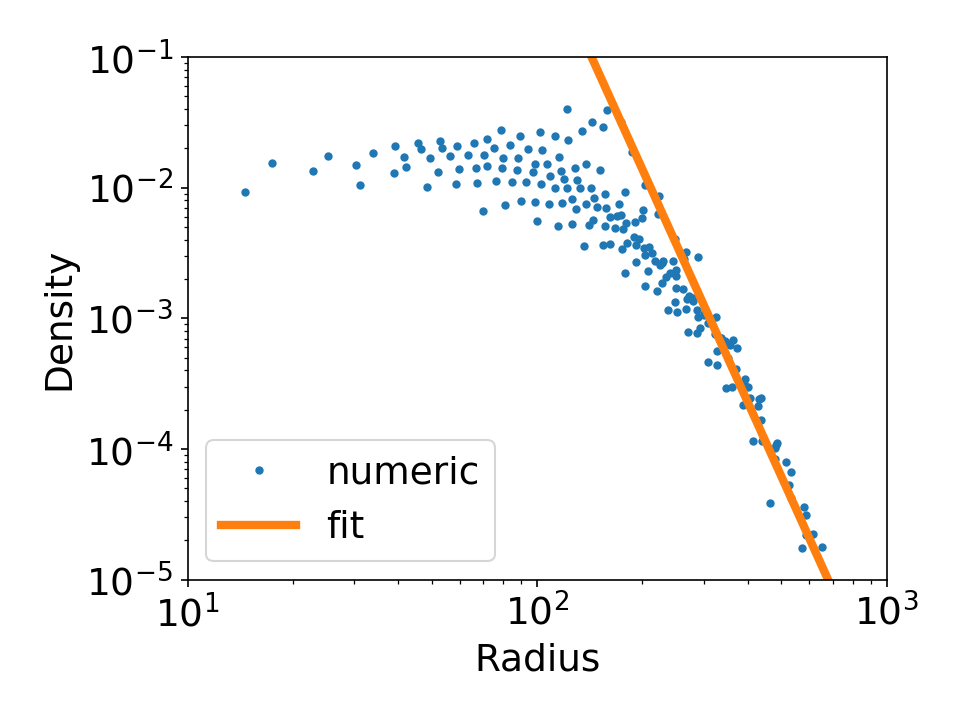}
    \caption{Radial density profile for a head on PBH collision, obtained from numerical simulations. The radius is given in units of the impactor radius, and the density in units of impactor (and target) density. The slope of the power law fit is -5.9, whereas the analytic prediction is -6. The numerical simulation confirms the analytic prediction.}
    \label{fig:pbh_density}
\end{figure*}

\begin{figure*}
\begin{tabular}{cc}
\subfloat{\includegraphics[width = 3.5in]{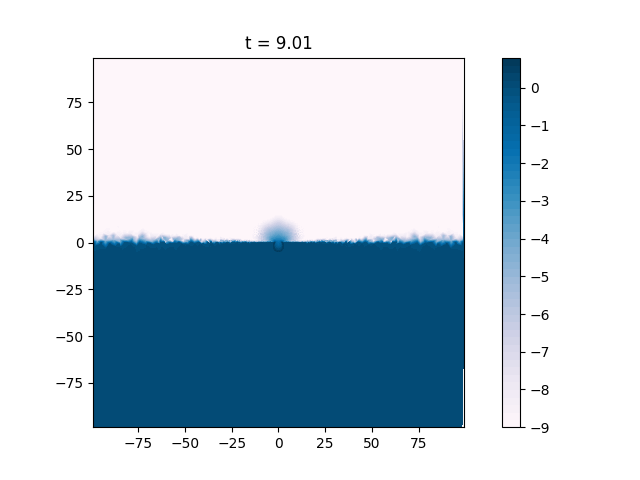}} &
\subfloat{\includegraphics[width = 3.5in]{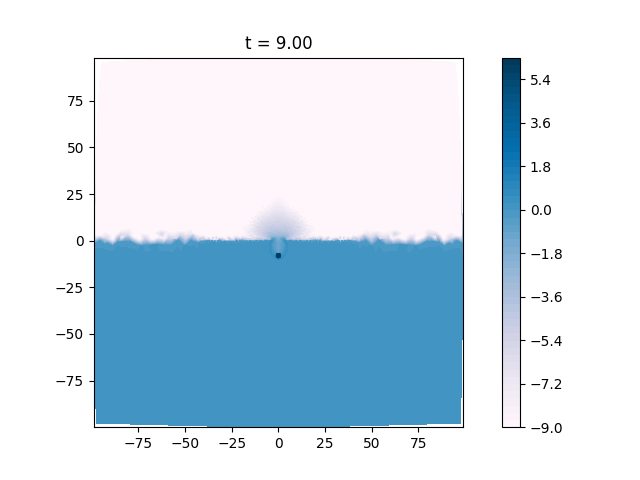}}\\
\subfloat{\includegraphics[width = 3.5in]{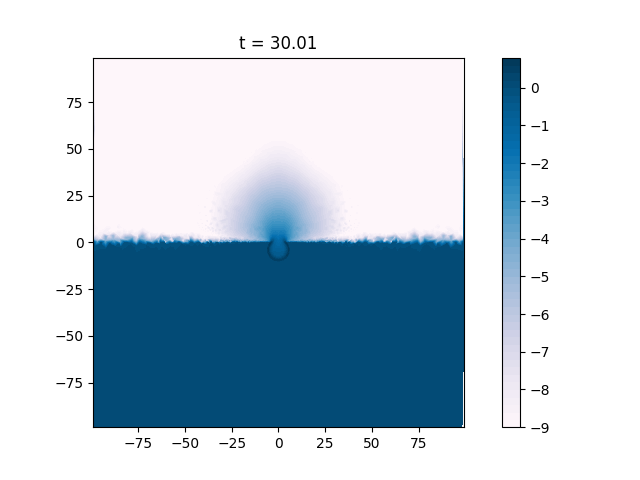}} &
\subfloat{\includegraphics[width = 3.5in]{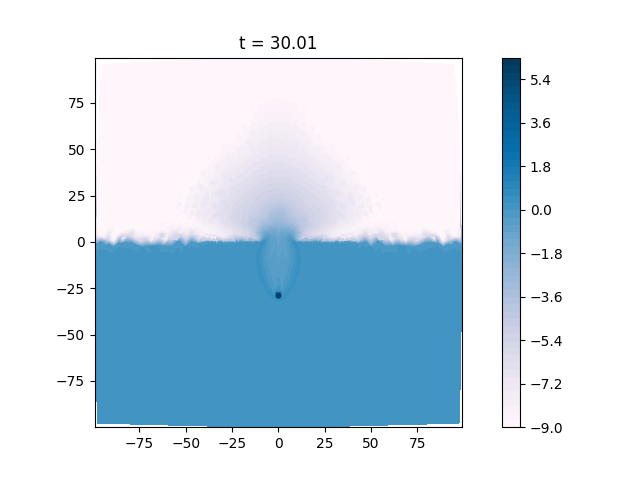}}\\
\subfloat{\includegraphics[width = 3.5in]{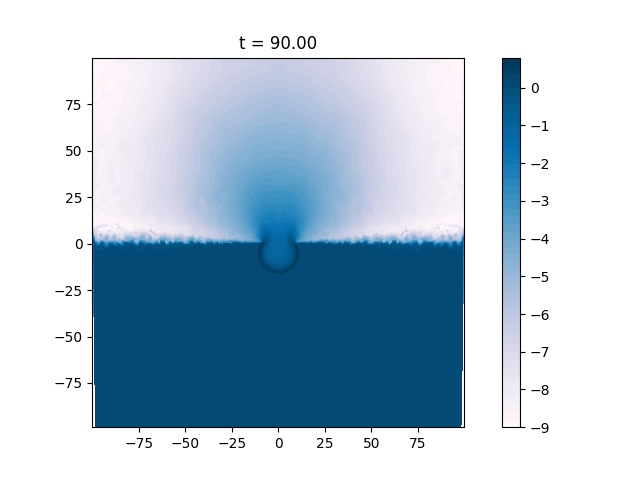}} &
\subfloat{\includegraphics[width = 3.5in]{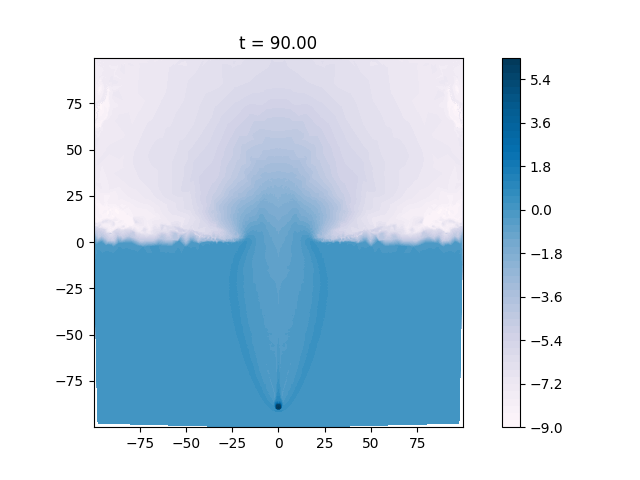}}
\end{tabular}
\caption{Log density snapshots from numerical simulations of a regular, head on impact (left) and an impact of a primordial black hole (right), both without gravity. Both impactors are the same size and initially move at the same velocity. The density of the regular impactor is the same as that of the ground, while the density of the PBH is six orders of magnitude higher. The distances shown in the images are normalised by the radius of the impactor. The time is normalised by the time it takes the impactor to traverse a distance equal to its radius. The density is normalised to the density of the target. These snapshots show that while the regular crater is decelerated quickly and deposits its energy in the target, the PBH keeps ploughing through. It is also apparent that the distribution of ejecta is different.}
\label{fig:sim_snapshot}
\end{figure*}

The second simulation is intended to describe a head on collision of a primordial black hole with the surface of the moon. The setup for this simulation is the same that described in the previous paragraph, except the density of the impactor is $10^6$ times greater than that of the target, and also we increase the resolution along the trajectory of the impactor inside the target, so the total number of points is larger (120,000). The radial density profile from this simulation is shown in figure \ref{fig:pbh_density}. From this figure we infer the slope $d \ln \rho / d \ln r = -5.9$. Our analytic theory predicts that $\beta = 1/3$, and so the slope of the density profile should be $d \ln \rho / d \ln r = -6$. The numeric simulations confirm our analytic prediction, and show that the behaviour of a PBH collisions are markedly different from a normal cratering event. In figure \ref{fig:sim_snapshot} we present sequential log density snapshots from both simulations to highlight the differences between them.

\section{Conclusions} \label{sec:conclusion}

In this work we considered the possibility that a primordial black hole collided with the moon. We show that a crater resulting from such a collision will have a different shape from a regular meteor impact, namely, a compact object impact crater will have steeper ejecta blanket. Most craters on the moon have ejecta blankets whose shape is compatible with a meteor impact. However, there are craters with steeper ejecta blankets whose origin does not involve exotic objects \citep[e.g. Mare Imbrium, see][]{Zhu2017EffectsDistribution}.

Steep ejecta blankets can therefore be used to flag craters of interest, but are not indisputable evidence of a compact object impact. A stronger evidence for black hole impact is the presence of a high pressure, pyrite-type phase of silica \citep{Kuwayama2005Geochemistry:Silica}. Regular impacts on the moon only produce the stishovite and seifertite phases of silica. This is because the new phase of silica appears above 40 TPa, whereas peaks pressure in a typical collision is around 1 TPa \citep{Kayama2018LunarSilica}. We therefore propose that steep ejecta blankets can be used to flag craters of interest, but to confirm a black hole impact origin would require finding the high pressure phase of silica.

If such a crater did exist on the moon, there is a possibility that it will have eroded away, due to meteor impacts or solar wind sputtering. For a one metre crater, the erosion time is over a billion years \citep{McDonnell1977TheSurface}, and for larger craters take even longer. Hence, we conclude that erosion is not a concern.

The smallest black holes that produce a detectable one metre crater have a mass of about $10^{19}$ g. We estimated that even for these small black holes, the impact rate is so low that, even if all dark matter is composed of such black holes, there's only about a 10\% chance that one collided with the moon throughout its lifetime. With such a low rate we are not able to place constraints on the dark matter distribution. However, the same considerations used here for the moon could apply just as well to any rocky body without an atmosphere. Therefore, with future missions to other rocky bodies in our solar system it will eventually be possible to use the method outlined in this work to constrain the mass distribution of primordial black holes.

\section*{Acknowledgements}

AY is supported by the Natural
Sciences and Engineering Research Council of Canada (NSERC), funding
reference \#CITA 490888-16. This work made use of the sympy \citep{Meurer2017SymPy:Python}, numpy \citep{Oliphant2006ANumPy} and matplotlib \citep{Hunter2007Matplotlib:Environment} python packages.

%%%%%%%%%%%%%%%%%%%%%%%%%%%%%%%%%%%%%%%%%%%%%%%%%%
\section*{Data Availability}

The source code for the numerical simulation and documentation can be found on github at \url{https://github.com/bolverk/huji-rich}.



\bibliographystyle{mnras}
\bibliography{main} % if your bibtex file is called example.bib

%\begin{thebibliography}{99}
%\end{thebibliography}

%%%%%%%%%%%%%%%%%%%%%%%%%%%%%%%%%%%%%%%%%%%%%%%%%%

%%%%%%%%%%%%%%%%% APPENDICES %%%%%%%%%%%%%%%%%%%%%

%\appendix



%%%%%%%%%%%%%%%%%%%%%%%%%%%%%%%%%%%%%%%%%%%%%%%%%%

% Don't change these lines
\bsp	% typesetting comment
\label{lastpage}
\end{document}